\documentclass[sigconf]{acmart}

\setcopyright{none}
\acmConference[ICCAD'26]{International Conference on Computer-Aided Design}{November 8--12, 2026}{San Jose, CA, USA}

\usepackage{makecell}
\usepackage{svg}
\usepackage{bm}
\usepackage{tikz}
\usepackage{algorithm}
\usepackage[noend]{algpseudocode}
\usepackage{afterpage}
\usepackage{enumitem}
\usepackage{multirow}
\usepackage{multicol}
\usepackage{graphicx}
\usepackage{hyperref} 

\usepackage{lipsum}
\usepackage{tabularx}
\usepackage{array}
\usepackage{booktabs}
\usepackage{listings}
\usepackage{xcolor}
\usepackage{framed}
\usepackage{svg}
\usepackage{py}
\usepackage{rotating}
\usepackage{amsmath}
\usepackage{bm}
\usepackage{tikz}
\usepackage{algorithm}
\usepackage{algpseudocode}
\usepackage{amsmath}

\newcommand{\cw}{\columnwidth}
\newcommand{\tw}{\textwidth}
\newcommand{\ctxcaptionsize}{\small}

\newcommand{\TT}[1]{\texttt{#1}}
\newcommand{\BF}[1]{\textbf{#1}}
\newcommand{\IT}[1]{\textit{#1}}

\newcommand{\mmsq}{mm\tsup{2}}
\newcommand{\umsq}{\textmu{}m\tsup{2}}

\newcommand{\tsup}[1]{\textsuperscript{#1}}

\AtBeginDocument{%
  }

\copyrightyear{2026}
\acmYear{2026}
\setcopyright{cc}
\setcctype{by}
\acmConference[ICCAD '26]{IEEE/ACM International Conference on Computer-Aided Design}{November 08--12, 2026}{San Jose, CA, USA}
\acmBooktitle{IEEE/ACM International Conference on Computer-Aided Design (ICCAD '26), November 08--12, 2026, San Jose, CA, USA}
\acmDOI{10.1145/3831252.3833953}
\acmISBN{979-8-4007-2873-0/2026/11}

\begin{document}

\title{Chipstitch: A Scalable and Cost-Efficient Platform for Accelerated Silicon Prototyping}

\author
{%
  Jeongeun Kim, Sabrina Yarzada, Paul Yi-Chia Chen, Yuanhan Li, Wenbo Zhu, and Christopher Torng
}

\affiliation{%
  \institution{Department of Electrical and Computer Engineering}
  \institution{University of Southern California}
  \city{Los Angeles}
  \state{CA}
  \country{USA}
}




\begin{abstract}
Silicon evidence serves as a critical receipt of credibility across academia and industry, validating innovation and signaling feasibility in ways that simulations alone cannot. Its value is difficult to quantify but widely recognized for driving hardware-software innovation. However, access to this receipt has remained severely constrained by high dollar costs and engineering effort barriers, which lock out innovators unable to afford it. To bridge this gap, we consider a new design paradigm called \IT{silicon query architectures} that spans EDA, architecture, and VLSI and aims to transform single-die area into a substrate for scalable, tessellated, intra-die silicon receipts, dropping the cost of entry by one to two orders of magnitude without sacrificing research-grade specifications. We concretely propose Chipstitch, a silicon query architecture realized with three contributions: an algorithmic three-layer approach to repeatable and scalable tessellation with a keystone algorithm, a scalable architecture-VLSI system interconnect that fits within interstitial regions of the tessellation with research-grade timing and clocking, and a grid-aligned perimeter-based power shutdown technique that mitigates overheads for research-grade power measurement. Using a vertically integrated methodology extending down to post-layout VLSI implementation, we demonstrate Chipstitch on a 25-site heterogeneous design scenario, achieving $9.9\times$--$14.5\times$ total cost advantage and up to two orders of magnitude reduction in single-receipt entry cost compared to a traditional multi-project wafer baseline in Skywater 130nm and Intel 16nm process nodes.
\end{abstract}

\maketitle

\section{Introduction}
\label{sec-intro}


\begin{figure*}[!t]
  \centering
  \includegraphics[width=\textwidth]{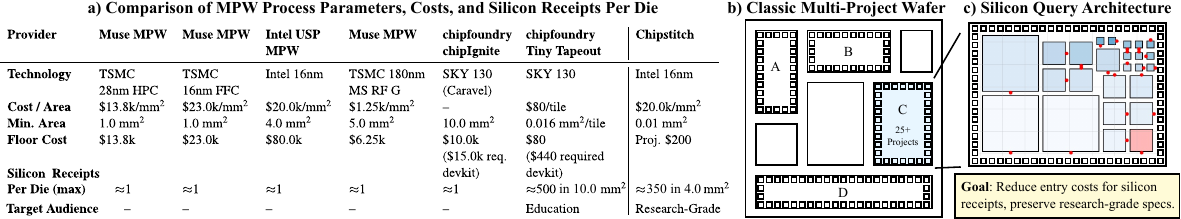}
  \vspace{-0.20in}
  \caption{Overview of low-volume silicon prototyping. (a)~Commercial MPW costs across technologies~\cite{muse-semi-web, mosis-web}, including education-first, non-research-grade options~\cite{efabless-caravel-web, efabless-tinytapeout-web, chipfoundry-web}; (b)~Classic MPW~\cite{kahng-mpw-ispd2004, conway-mosis-mpc-1981, tomovich-mosis-1988, cohen-mosis-1981};
  (c)~Scalable intra-die silicon receipts, preserving research-grade specifications (e.g., high-speed GHz clocking, isolated power measurement), verified to post-layout.}
  \label{fig-intro}
  \vspace{-0.15in}
\end{figure*}

Silicon evidence is a well-recognized artifact of progress in both industry and academic literature. Landmark systems such as Eyeriss~\cite{chen-eyeriss-jssc2017} and large-scale industrial accelerators such as Google's TPU~\cite{jouppi-google-tpu-isca2017,jouppi-tpuv4i-isca2021} derive much of their impact not only from architectural novelty, but from the credibility conferred by fabricated hardware. Beyond verification and performance validation, it serves a broader and often implicit role: silicon evidence acts as a \emph{receipt of credibility between two parties}. The value of this receipt is difficult to quantify but widely recognized, for example: between a government funding agency and its performers, between investors and startup founders, between academic reviewers and paper authors. Further, many microelectronics funding agencies~\cite{darpa-eri-web} require performers to produce silicon, despite the concluding findings not differing significantly from detailed pre-silicon simulations. In each case, silicon evidence signals feasibility and reduces uncertainty in a way that models and simulations alone cannot.
However, access to this receipt has remained heavily constrained by high dollar cost and engineering effort barriers, which lock out innovators unable to afford it.

For over four decades, multi-project wafer (MPW) services have defined these barriers for low-volume manufacturing by amortizing fabrication costs across projects on a single wafer~\cite{cohen-mosis-1981, conway-mosis-mpc-1981, tomovich-mosis-1988}. However, they still impose high minimum-entry fees and coarse granularities (i.e., a large minimum-area purchase) that are only accessible by the very few~\cite{mosis-web, muse-semi-web}.
\IT{Dollar cost} is dominated by fabrication costs and total compensation for a team of chip design engineers. \IT{Engineering effort} manifests through a full-team chip development effort over a project span of six months up to a few years to derisk the delivery of successful silicon.
Figure~\ref{fig-intro} shows a classic MPW with multiple projects A to D, each representing a physically independent die, aggregated on a single wafer. Figure~\ref{fig-intro} also collects table data comparing commercial MPW service offerings across technology nodes (180nm--16nm), listing how the cost per \mmsq{} and minimum area requirements combine into high floor costs of entry for hardware-software systems innovation. This is particularly true in FinFET-class nodes most meaningful for innovation, with minimum-area seats of $1\times1 mm^2$ to $2\times2 mm^2$ in 12nm-16nm technologies quoted at \$20K--\$80K costs of entry~\cite{mosis-web, muse-semi-web}.
For individual chip design innovators, the fundamental underlying mechanism is a realistic \emph{economic} evaluation of the dollar cost and engineering effort for silicon against the value of the credibility and the engineering insights gained. For the broader field, high costs of entry restrict a field's ability to innovate. As a concrete example, one recent circuits advance in advanced cryptographic hardware requires as small as 0.076\mmsq{} of GF~12nm area~\cite{sreedhar-xgcd-esserc2024}, but Figure~\ref{fig-intro} shows that the minimum area would cost over 10$\times$ more in 16nm- and 12nm-class MPW services. Such projects only survive by attaching to larger projects riding on the same shuttle~\cite{koul-onyx-vlsi2024}, locking out most researchers from critical silicon resources.

This work proposes that these long-standing cost floors are not fundamental limits, and that a design paradigm we call \IT{silicon query architectures} can transform single-die area into a substrate for large numbers of research-grade intra-die silicon receipts that are tessellated together (i.e., tetris), thereby amortizing costs and effort down to one or more orders of magnitude below that of traditional MPWs, while maintaining research-grade specifications (Figure~\ref{fig-intro}c). There is a way to measure the per-site result as if the entire chip contains only that site. Unlike an MPW, in which projects are physically separate and diced apart (e.g., Figure~\ref{fig-intro}b), intra-die sites logically connect to each other and to shared resources (e.g., IO pads, clock generators). Fabricated chips are provided to each project owner or shared for reuse.
Qualitatively, such a methodology has two important results: (1)~intra-die silicon-proportional costs, rather than reserving a full chip, and (2)~a requirement for only block-level design work, leaving full-chip delivery to be amortized by on-die aggregator services, similar to how MPW services operate today.

Proper exploration of the implementation and merits of such a chip design methodology are beyond the scope of a single work, and we instead focus on addressing the most critical EDA/CAD bottlenecks that unlock the physical realization in the first place, leaving future works to explore further.
The key challenges revolve around the repeatable, scalable, and efficient tessellation of many sites into a single die around an active, logical, on-die interconnect.

Silicon query architectures have three major unaddressed challenges that contrast with the capabilities of prior literature: (1)~large tessellations create a physical manufacturability challenge (i.e., DRC) that can rival that of a commercial SoC, (2)~each site has unbounded timing assumptions (e.g., per-site clock tree insertion delays) that require unbounded area to meet timing at the system interface, thereby overflowing the fixed interstitial area between sites in the tessellation, (3)~the idle power of all sites on the chip can mask the sensitive power measurement for a target site.

This work proposes Chipstitch, a silicon query architecture we realized to directly address the three challenges.
First, we develop an \BF{algorithmic three-layer approach to repeatable and scalable tessellation}. The fundamental layer is the discretization of physical design into site templates, which narrows the set of manifesting DRC classes to a small discrete set that appears identically in a repeated manner across the tessellation due to spatial isomorphism, and can then be systematically fixed at scale. The second layer proposes a \IT{keystone algorithm} that transforms the physical integration problem into a geometric variant of set cover (therefore NP-complete) that produces a layout manifesting each unique DRC scenario once. A clean keystone layout thereby provides a certificate for the elimination of entire classes of DRC across a combinatorial space of tessellations, and need only be executed once per template library set and per technology. The third layer observes that one can then deploy algorithmic methods for tessellation (e.g., bin packing) to shrink bounding box areas, which are valid and feasible as long as the prior layers are true (i.e., discretization, keystone certificate). Our approach allows heterogeneous sites to be algorithmically tessellated into any of a combinatorial space of potential layouts (i.e., tetris) with correct-by-construction physical integration benefits. Without this foundation, new tessellations are as physically difficult to build as a commercial SoC.
Second, we design a \BF{scalable, hybrid architecture-VLSI system interconnect} that fits within the fixed interstitial region between sites in the tessellation. We address unbounded timing, clocking, and other challenges with fixed-area solutions. This enables scalable expansion to large site counts with research-grade timing and clocking.
Third, we propose \BF{grid-aligned, perimeter-based power domains} that allow per-site power shutoff of idle power for inactive sites, allowing for research-grade power measurement of the active site without requiring users to have low-power design expertise.


\begin{figure*}[t]

  \centering
  \includegraphics[width=0.90\tw]{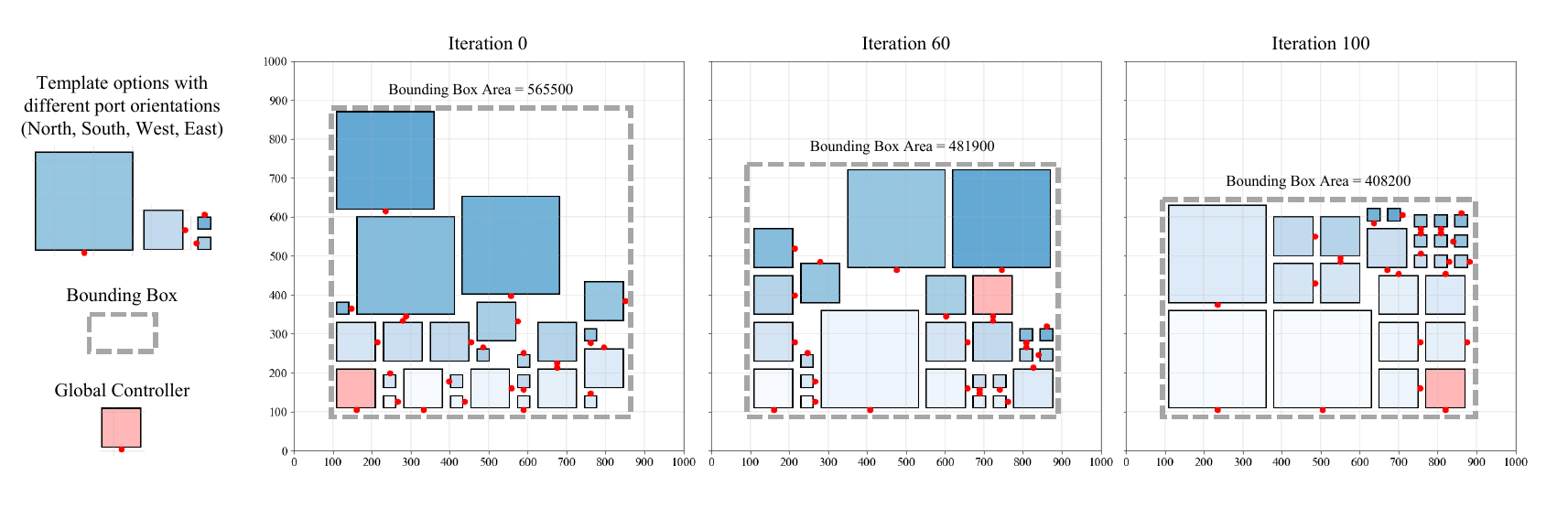}
  \vspace{-0.2in}
  \caption{Stochastic solver for tessellation operating on a templated, grid-based layout canvas.
  Sites iteratively packed into smaller bounding boxes. Logical start is red (global controller). Blocks colored darker in connection sequence. Red dots =~ports.}

  \label{fig-stochastic-iterations}
  \vspace{-0.15in}

\end{figure*}

Our results show these techniques eliminate multiple integer factors of area overhead, achieving for one specific studied scenario a $9.9\times$--$14.5\times$ total cost advantage over a traditional MPW baseline, and up to two orders of magnitude lower single-receipt entry cost in Skywater 130nm and Intel 16nm. We contribute the following:

\begin{itemize}[noitemsep, topsep=0pt]
\item We identify silicon query architectures as a distinct design paradigm spanning EDA, architecture, and VLSI, requiring new advances to achieve a repeatable and scalable substrate for aggregating intra-die silicon as research-grade receipts of credibility between two parties.
\item We develop an algorithmic three-layer approach to repeatable, scalable tessellation with physical design discretization, keystone algorithm, and automated packing methods.
\item We design a scalable, hybrid architecture-VLSI system interconnect that decouples area overheads from site count and complexity to fit within fixed interstitial area in a tessellation with research-grade clock speeds and timing.
\item We propose grid-aligned perimeter-based power shutdown domains to enable research-grade power measurement, without requiring low-power VLSI expertise for users.
\end{itemize}

To the best of our knowledge, we present the first integrated EDA, architecture, and VLSI solution that breaks past traditional multi-project wafer boundaries by transforming single-die area into a substrate for scalable intra-die silicon receipts, reducing single-receipt entry costs significantly without sacrificing research-grade specifications. We hope our framework serves as a stepping stone toward accessible silicon and a future of lowered barriers to hardware-software innovation.


\section{Silicon Query Concepts and Experiments}
\label{sec-system}

We define key concepts to bridge real-world designs as silicon receipts to the algorithms abstraction presented in the next section.

We previously introduced the top-down intuition of a silicon query architecture, but the bridge is best clarified from bottom up. We first define a primitive \BF{silicon query} as the pairing of a design datapath and the active data flowing through it, specifically scoped for the purpose of isolated silicon measurement. Intuitively, it is the portion of the hardware and its measurement waveform whose numbers answer a question (hence silicon query, for performance, energy, etc.). A \BF{silicon query site} is the physical location on-chip where the execution and measurement take place. This is important because we must delineate measurable regions of a chip, with the correct entities in the intended boxes. Each rectilinear shape in Figure~\ref{fig-intro}c corresponds to a silicon query site (henceforth simply called a site), and physical tessellation deals with this abstraction boundary. Finally, a \BF{silicon query architecture} (SQA) defines how one or more silicon query sites are integrated together to share die resources. This includes logical aspects (e.g., pads, configuration and data delivery interconnects, per-site power controls, resets) and physical aspects (e.g., packing, port locations, shapes). As a result, a tessellation with 25 heterogeneous sites (i.e., tetris pieces) can be configured sequentially to answer 25 sets of questions independently for each silicon receipt measurement.

We \BF{explicitly scope our work to a restricted class of silicon receipts with an intra-site, stage-execute-readback design style} for post-fabrication experiments: those that stage data slowly from off-chip into the user site, run the measured experimental trace entirely on-chip at full-silicon speeds, then read back data slowly off-chip. This pattern is common in many academic circuits literature~\cite{feng-amber-jssc2024, koul-onyx-vlsi2024}. It captures a restricted but still useful class of silicon receipts of credibility. Measured experiments run entirely intra-site. Only one site is activated and isolated (others power-gated).
Importantly, it provides (1)~the per-site independence to reason about the system-level problems, algorithms, and proofs that we present in this work; and (2)~design benefits. For example, intra-site measurement only occurs in an execute phase, so the outer system has no performance requirement during staging and readback. We can therefore design a low-bandwidth interconnect into the sensitive fixed-area interstitial regions of the tessellation.
All off-chip and system performance overheads (e.g., JTAG, DRAM, Chipstitch interconnect) are explicitly not participants in the measured experiments.
We leave other classes to future researchers.


\section{Algorithms for Scalable Tessellation}
\label{sec-algorithms}

We discuss an algorithmic approach to enable tessellation and physical manufacturability repeatably and at scale.

\subsection{Grid-Based Templatization}

\begin{table*}[t]
\centering
\caption{Stochastic placement and routing results.
Rows correspond to Template Diversity (Y-axis) and columns correspond to
Total Site Count (X-axis). Each case reports Solver Time (s), Track/Bbox (\%),
and Utilization (\%).}
\label{tbl-stochastic-matrix}
\vspace{-0.1in}

\footnotesize
\setlength{\tabcolsep}{4pt}
\scriptsize
\setlength{\tabcolsep}{4pt}

\begin{tabular}{c cccc cccc cccc cccc cccc}
\toprule

\multirow{2}{*}{\centering\shortstack{\# of Sites\\vs\\Diversity}}

& \multicolumn{4}{c}{\textbf{5 sites}}
& \multicolumn{4}{c}{\textbf{10 sites}}
& \multicolumn{4}{c}{\textbf{20 sites}}
& \multicolumn{4}{c}{\textbf{50 sites}}
& \multicolumn{4}{c}{\textbf{100 sites}} \\

\cmidrule(lr){2-5}
\cmidrule(lr){6-9}
\cmidrule(lr){10-13}
\cmidrule(lr){14-17}
\cmidrule(lr){18-21}

& ID & T(s) & T/B(\%) & U(\%)
& ID & T(s) & T/B(\%) & U(\%)
& ID & T(s) & T/B(\%) & U(\%)
& ID & T(s) & T/B(\%) & U(\%)
& ID & T(s) & T/B(\%) & U(\%) \\

\midrule

1 (homogeneous)
& P1  & 75.41    & 0.90 & \textbf{98.59}
& P6  & 112.68   & 0.91 & \textbf{94.23}
& P11 & 149.73   & 0.97 & \textbf{83.97}
& P16 & 261.53   & 0.59 & \textbf{86.35}
& P21 & 1008.95  & 0.09 & \textbf{90.83} \\

2
& P2  & 157.24   & 1.17 & \textbf{89.22}
& P7  & 315.15   & 0.96 & \textbf{87.40}
& P12 & 284.15   & 0.88 & \textbf{86.50}
& P17 & 709.19   & 0.63 & \textbf{89.65}
& P22 & 3576.93  & 0.22 & \textbf{94.84} \\

3
& P3  & 429.50    & 1.32 & \textbf{89.64}
& P8  & 661.82    & 1.27 & \textbf{84.62}
& P13 & 1012.55   & 0.73 & \textbf{89.67}
& P18 & 1915.38   & 0.62 & \textbf{88.99}
& P23 & 11394.33  & 0.19 & \textbf{89.91} \\

4
& P4  & 1945.20   & 1.26 & \textbf{83.77}
& P9  & 2867.13   & 1.58 & \textbf{87.92}
& P14 & 4929.31   & 1.35 & \textbf{86.06}
& P19 & 8736.11   & 0.72 & \textbf{84.11}
& P24 & 48085.70  & 0.29 & \textbf{93.90} \\

5 (mixed)
& P5  & 11526.87   & 1.92 & \textbf{85.07}
& P10 & 11112.95   & 1.80 & \textbf{81.13}
& P15 & 23196.05   & 1.18 & \textbf{86.70}
& P20 & 44600.68   & 0.67 & \textbf{75.99}
& P25 & 202363.78  & 0.41 & \textbf{87.74} \\

\bottomrule
\end{tabular}
\vspace{-0.15in}
\end{table*}

The FinFET and more advanced technology nodes have large minimum areas (Figure~\ref{fig-intro}), but larger tessellations leads to a combinatorial explosion in DRC interactions due to fine-grained inter-site coupling and system logic. As site count increases, verification cost grows with the number of possible geometric adjacencies, making direct application of classical packing methods insufficient.

\begin{figure}[!t]
  \centering
  \includegraphics[width=0.92\cw]{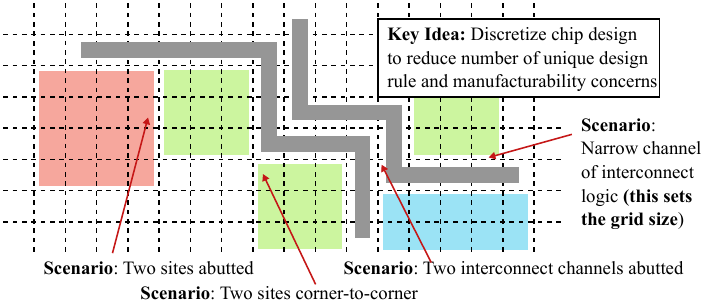}
  \vspace{-0.15in}
  \caption{Intuition underlying discretization of physical design into a template library for grid-based tessellation.}
  \label{fig-algo-grid}
  \vspace{-0.1in}
\end{figure}

To bound this growth, we discretize the layout space using grid-based templatization. Silicon query sites are restricted to a small library of certified templates instantiated on a common grid, each with pre-validated internal DRC compliance and standardized boundary interfaces. This reduces possible placement interactions to a finite, enumerable set of template-to-template and template-to-interstitial scenarios.
Rather than verifying each layout instance, we pre-validate this bounded set of interactions once. As a result, DRC complexity depends only on template diversity and grid granularity, rather than total site count.
The grid is defined per technology node so that a single interstitial channel satisfies all routing and DRC constraints. Under this formulation, tessellation reduces to packing over certified primitives, enabling standard optimization techniques while preserving manufacturability guarantees (see Figure~\ref{fig-algo-grid}). This discretization transforms layout verification from an open-ended geometric problem into a bounded combinatorial optimization problem, and introduces the complementary challenge of fixed-area interconnect design addressed in Section~\ref{sec-architecture}.

\subsection{Stochastic Packing under Constraints}
\label{sec-stochastic}

Figure~\ref{fig-stochastic-iterations} illustrates representative iterations of our solver operating upon this grid, which combines simulated annealing with skyline-based heuristics to pack 25~sites drawn from three template types into a progressively shrinking bounding region. With manufacturability constraints bounded through grid-based templatization, site tessellation reduces to a combinatorial packing problem over certified primitives. This enables the use of stochastic optimization techniques for automated floorplanning. Because all sites are instantiated on a common grid and interstitial channels occupy fixed-width tracks, the solver operates entirely within a pre-verified configuration space. The active on-die interconnect is constrained to route through interstitial grid channels and to abut each site at least once to ensure full connectivity.

Table~\ref{tbl-stochastic-matrix} shows 25 packing scenarios ranging from 5--100 sites, including both homogeneous and heterogeneous template mixes. Across experiments, silicon utilization ranges from 76\% to 98\%, approaching the ideal case in which sites and interstitial tracks fully occupy the die. Importantly, integration overhead remains bounded rather than growing superlinearly as in conventional aggregation.


\subsection{Keystone Layout Methodology}
\label{sec-drc-algorithm}

\begin{figure}[!t]
  \centering
  \vspace{-0.05in}
  \includegraphics[width=1.03\cw]{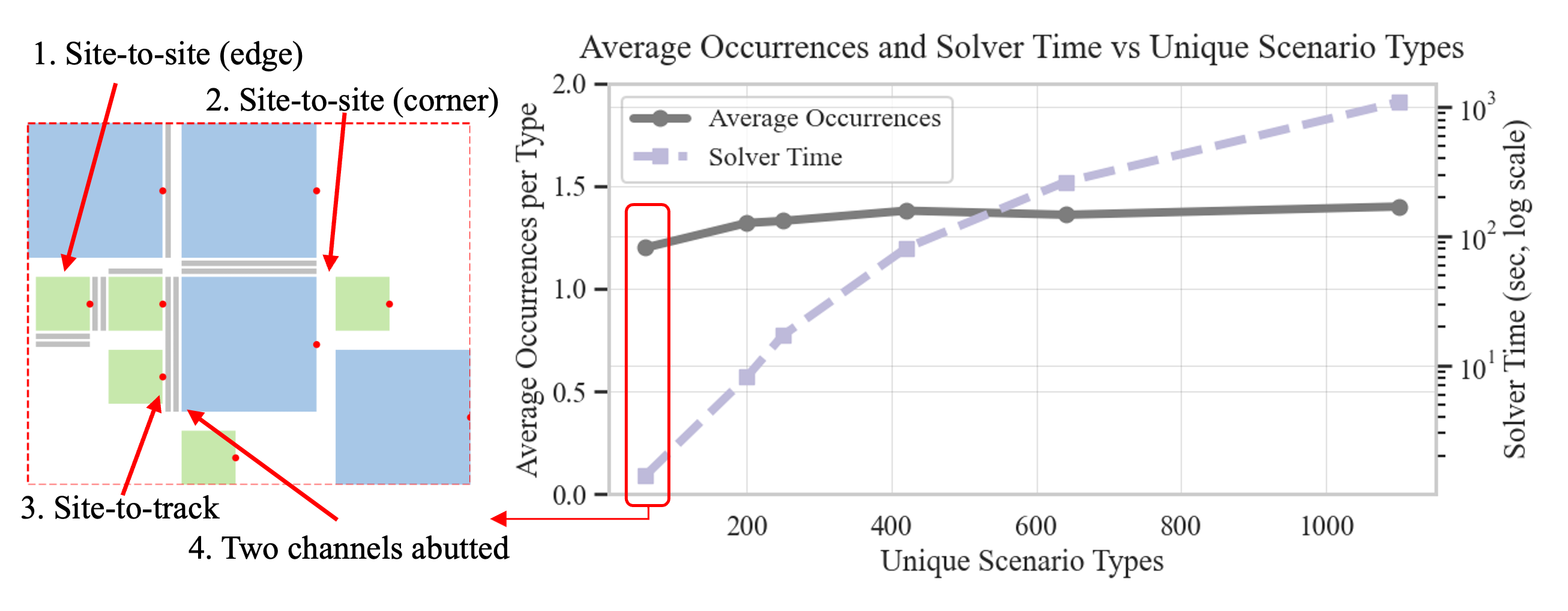}
  \vspace{-0.35in}
  \caption{Evaluation of Keystone DRC enumeration algorithm. Left: Keystone floorplan (2 templates).
  Right: Avg scenario count and solver time vs.~ num unique scenario types.}

  \label{fig-drc-enumeration}
  \vspace{-0.15in}
\end{figure}

The prior subsection proceeded with packing but left unanswered how templates are designed to avoid integration DRC challenges. The highlight of our approach is illustrated in Figure~\ref{fig-drc-enumeration}, in the concept of the \emph{keystone layout} and the algorithm used to construct it. The keystone layout is a single unified layout instance that embeds all manufacturability-critical adjacency configurations induced by the discretized site template system. The goal in this section is to construct a layout that passes full-chip DRC under production rule decks, such that any composition of the same templates is guaranteed to be DRC-clean.

A conventional approach performs DRC validation at instance level. Each new integration requires re-verification under full-chip context, as global spacing and coupling effects cannot be inferred from isolated patterns. As integration scale increases, the number of distinct adjacency configurations grows combinatorially, leading to repeated and increasingly expensive DRC closure.
Under grid-based templatization, however, placement interactions are restricted to a finite and enumerable set of template-to-template and template-to-interstitial scenarios. This bounded interaction space enables a different approach: instead of verifying each instance, we construct a single layout that enumerates all discrete scenarios.


We formulate keystone construction as a constrained coverage problem. The objective is to embed every DRC-sensitive configuration at least once while minimizing layout area. This problem is equivalent to a geometric variant of set cover and is therefore NP-complete. After discretization, 
for $G$ denoting the set of discrete grid locations, $C$ the set of component variants,
$U$ the number of unique scenario types, and $k$ average number of candidate placements,
the total number of feasible placements is $|\mathcal{P}| = O(|G|\cdot|\mathcal{C}|)$, with an exponential search space of size $O(2^{|\mathcal{P}|})$. Despite this worst-case complexity, we exploit structure in the discretized space to derive a practical greedy approach. By restricting search to scenario-relevant placements and exploiting locality, run time scales approximately as $O(U \cdot k \cdot P^2)$ in practice. The algorithm iteratively selects placements maximizing coverage of uncovered scenarios while controlling area growth and avoiding redundant embeddings, with controlled stochastic exploration to escape local optima.

As illustrated in Figure~\ref{alg:keystone-layout}, we construct the keystone layout using an $\epsilon$-greedy placement algorithm that iteratively realizes a predefined set of spatial scenarios $\Sigma$. Each scenario encodes a required geometric relation (e.g., abutment, corner adjacency, or site–track alignment), and the algorithm terminates once all scenarios are covered. At each iteration, candidate placements are restricted to component types appearing in the remaining uncovered scenarios $\Sigma_{\text{rem}}$, ensuring that every placement directly contributes to coverage. Each candidate placement $(c,x,y)$ is evaluated using

\vspace{-0.10in}

\[
f = \max\!\bigl(0,\ w_{\text{cov}}\,|\sigma_{\text{new}}| + w_{\text{den}}\,s_{\text{den}} - w_{\text{red}}\,n_{\text{red}}\bigr),
\]

\vspace{-0.00in}

where $|\sigma_{\text{new}}|$ counts newly covered scenarios, $s_{\text{den}}$ measures packing density, and $n_{\text{red}}$ penalizes redundant relations. Candidates not introducing new coverage are discarded. To avoid local optima, we select the highest-scoring move with probability $1-\epsilon$ and a random valid move with probability $\epsilon$. This formulation prioritizes coverage efficiency and compactness, yielding near-minimal layouts while ensuring complete realization of $\Sigma$. Keystone coverage can be systematically extended to additional rule classes by adding configurations to the template library set. Note that the keystone approach does not cover global signoff concerns such as density, fill, sealrings, and mask alignment cell insertion, but these are also not combinatorial problems.

Figure~\ref{fig-drc-enumeration} shows the scaling behavior of keystone construction. The average embedding multiplicity remains nearly constant as scenarios scale, indicating that layout size grows approximately linearly. For example, increasing the number of scenarios from 60 to 1100 results in an average embedding multiplicity of between 1.2 and 1.4 manifestations of each DRC scenario in the keystone layout. In an extreme case with 1100 scenarios across 20 templates, construction completes in 1096s, while a typical configuration of 200 scenarios and 8 templates completes in 8.22s. Once verified, the keystone layout serves as a manufacturability certificate for the entire template library. This shifts DRC complexity from per-instance verification to a one-time, bounded certification process, enabling scalable integration under continuous deployment settings.

Table~\ref{tab:keystone_drc} shows a microbenchmark studying whether fixes derived from the keystone transfer across three tessellations (T1--T3) constructed from the same template library. We consider three spacing-related DRC categories.
Each column in Table~\ref{tab:keystone_drc} shows how fixing the representative keystone violation eliminates the same violation across all three tessellations. For example, fixing four Keystone hotspots in the end-to-end spacing category eliminates all 11 hotspots of the same category across T1--T3, demonstrating one-to-many fix transfer in this simplified demonstration.


\section{Scalable Hybrid Architecture and VLSI Design}
\label{sec-architecture}

Intra-die sites logically connect to each other and to shared resources (e.g., IO pads, clock generators), but packing leaves only a narrow, fixed-width interstitial channel traversing sites within which to achieve research-grade timing. This creates a tightly constrained design problem requiring architecture-VLSI co-design. The primary constraint is bounded area, to fit within a fixed footprint independent of site count, frequency, or design complexity. Violating this breaks the packing solution and leads to DRC failure.

\begin{figure}[t]
\footnotesize
  \centering
  \begin{algorithmic}[1]
  \Procedure{KeystoneLayout}{sites, tracks, weights, $\epsilon$, max\_retries}
    \State $\Sigma \gets$ \TT{generate\_scenarios}(sites, tracks)
    \For{each retry attempt}
      \State Initialize layout with largest site at center
      \State $uncovered \gets \Sigma$
      \While{$uncovered \neq \emptyset$}
        \State $candidates \gets$ components appearing in $uncovered$
        \State $moves \gets \emptyset$
        \For{each placed $p$, candidate $c$}
          \For{each adjacent position of $p$}
            \If{position is overlap-free}
              \State $new \gets uncovered \cap \TT{relations}(c, \text{position})$
              \If{$new = \emptyset$} \textbf{continue} \EndIf
              \State $score \gets w_{cov}|new|$
              \State $score \mathrel{+}= w_{den}\cdot \TT{density}(c,\text{position})$
              \State $score \mathrel{-}= w_{red}\cdot \text{redundancy}(c,\text{position})$
              \State Add $(score, c, position, new)$ to $moves$
            \EndIf
          \EndFor
        \EndFor
        \If{$moves = \emptyset$}
          \State $best \gets$ \TT{forced\_placement}(uncovered)
        \ElsIf{$\mathrm{rand}() < \epsilon$}
          \State $best \gets$ random move from $moves$
        \Else
          \State $best \gets$ highest-scoring move
        \EndIf
        \State Place $best.component$ at $best.position$
        \State $uncovered \gets uncovered \setminus best.new$
      \EndWhile
      \If{$uncovered = \emptyset$} \textbf{break} \EndIf
    \EndFor
    \State \Return layout
  \EndProcedure
  \end{algorithmic}
  \vspace{-0.1in}
  \caption{
    Keystone layout generation via $\epsilon$-greedy placement.
    Algorithm covers spatial scenarios $\Sigma$ with three mechanisms:
    (1) \textbf{scenario-driven pruning} to restrict candidates to components in uncovered scenarios;
    (2) \textbf{redundancy-aware scoring} to prioritize placements maximizing new coverage, penalize duplication, and favor compact packing;
    (3) \textbf{$\epsilon$-greedy exploration with fallback} to balance greedy select with stochastic exploration, ensuring progress when no good move exists.
  }
  \label{alg:keystone-layout}
  \vspace{-0.1in}
\end{figure}

\begin{table}[t]
\setlength{\tabcolsep}{2pt}
\centering
\caption{Spatial isomorphism and combinatorial impact from keystone layout K.
DRC rules hidden for this technology.
}
\label{tab:keystone_drc}
\vspace{-0.1in}

\resizebox{\columnwidth}{!}{
\begin{tabular}{lccc|ccc|ccc}
\toprule

\textbf{Category}
& \multicolumn{3}{c|}{\textbf{End-to-end}}
& \multicolumn{3}{c|}{\textbf{Global/metal spacing}}
& \multicolumn{3}{c}{\textbf{Via spacing}} \\

\textbf{Rule Family}
& \textbf{M*\_03}
& \textbf{M*\_41}
& \textbf{M*\_44}
& \textbf{M*\_02}
& \textbf{M*\_81}
& \textbf{M*\_301}
& \textbf{V*\_22}
& \textbf{V*\_54}
& \textbf{V*\_128} \\

\midrule

\textbf{Keystone}
& 1 & 2 & 1
& 1 & 1 & 2
& 1 & 1 & 1 \\

\textbf{Tes. T1}
& 0 & 2 & 0
& 0 & 0 & 1
& 1 & 1 & 0 \\

\textbf{Tes. T2}
& 2 & 3 & 0
& 2 & 0 & 4
& 2 & 0 & 3 \\

\textbf{Tes. T3}
& 0 & 4 & 0
& 0 & 0 & 1
& 2 & 0 & 2 \\

\midrule

\textbf{After Fix (K)}
& 0 & 0 & 0
& 0 & 0 & 0
& 0 & 0 & 0 \\

\midrule

\textbf{(T1/T2/T3)}
& \multicolumn{3}{c|}{\textbf{0 / 0 / 0}}
& \multicolumn{3}{c|}{\textbf{0 / 0 / 0}}
& \multicolumn{3}{c}{\textbf{0 / 0 / 0}} \\

\bottomrule
\end{tabular}
}
\vspace{-0.15in}
\end{table}


\begin{figure*}[!t]
  \centering

  \begin{minipage}[c]{0.70\textwidth}
    \includegraphics[width=1.0\tw]{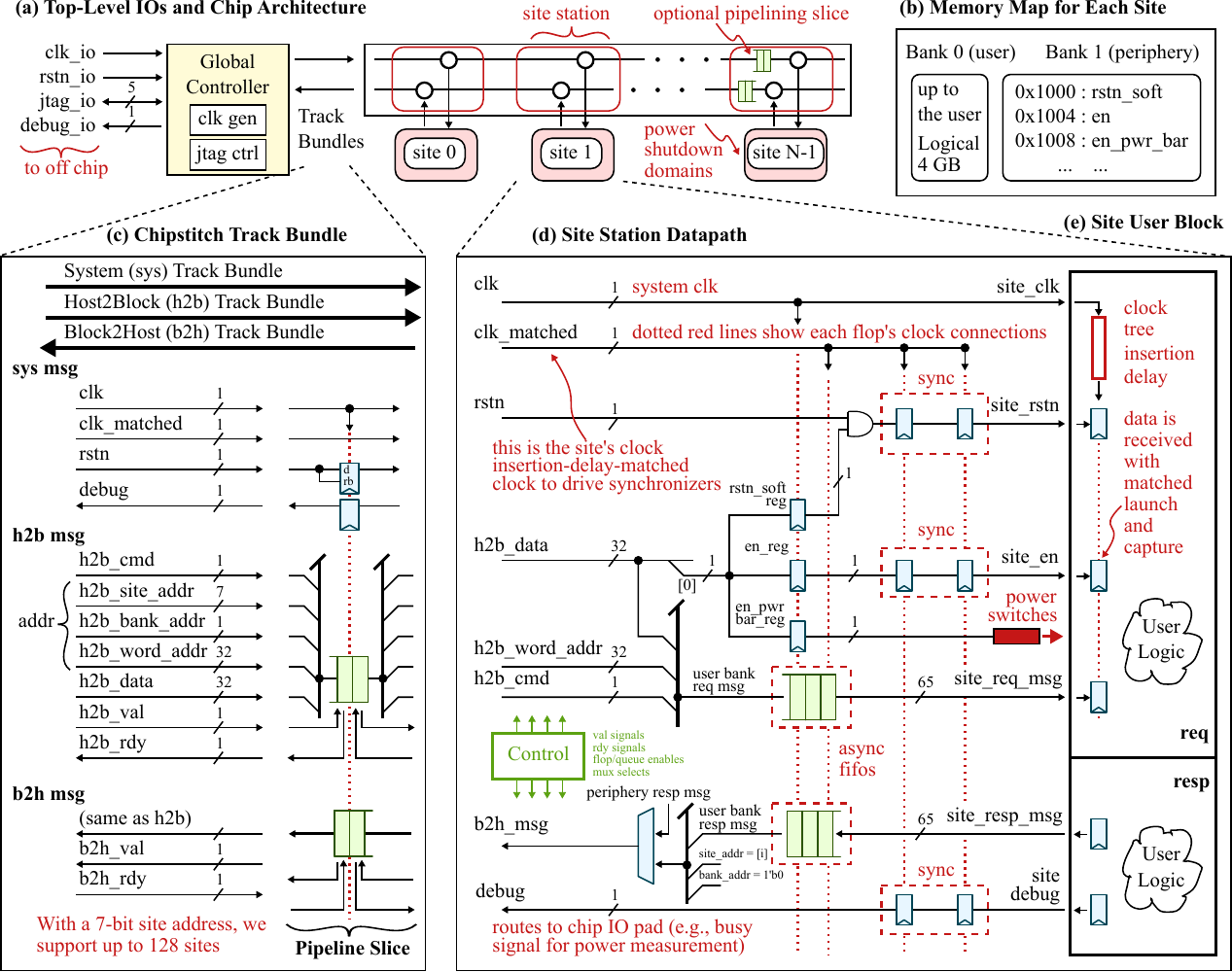}
  \end{minipage}\hfill%
  \begin{minipage}[c]{0.27\textwidth}
    \caption{Chipstitch Block Diagrams: (a)~top-level logical architecture including global controller, interconnect with per-site stations, and each site user block within a power shutdown domain; (b)~memory-mapped IO approach indexes into each site with a site address and allows accessing user and periphery banks; (c)~Chipstitch track bundle design shows message field bitwidths designed not to scale with site count~N, and showing pipeline slice logic for crossing physical distance; (d)~site station datapath with clock crossing details highlighted in dotted red vertical lines; (e)~site user block detailing the clocking challenge.}
    \label{fig-chipstitch-arch}
  \end{minipage}\hfill

  \vspace{-0.05in}
\end{figure*}

This constraint introduces several challenges at the architecture-VLSI boundary. First, timing closure becomes an area problem. While ASIC tools routinely fix timing by inserting buffers, such fixes increase area and can exceed the available routing tracks. Therefore, the design must be architected to avoid triggering tool-driven timing repair. Second, the fixed-area constraint directly limits interconnect bandwidth. High-throughput networks require wider datapaths and buffering, which are incompatible with the narrow channel budget. More critically, the packing solution effectively reduces available routing resources to a minimum spanning tree. This severely restricts topology choices: even low-bandwidth star networks become infeasible because all traffic must traverse a single physical channel, causing routing demand to scale with the number of sites.
These constraints motivate our hybrid architecture-VLSI design, which reshapes the interconnect to align with physical limits rather than optimizing for conventional performance metrics. Figure~\ref{fig-chipstitch-arch} shows the resulting design.

\subsection{Single-Master 1D Bi-Directional Mesh NoC}

\begin{figure}[t]
\centering
  \centering
  \includegraphics[width=1.0\cw]{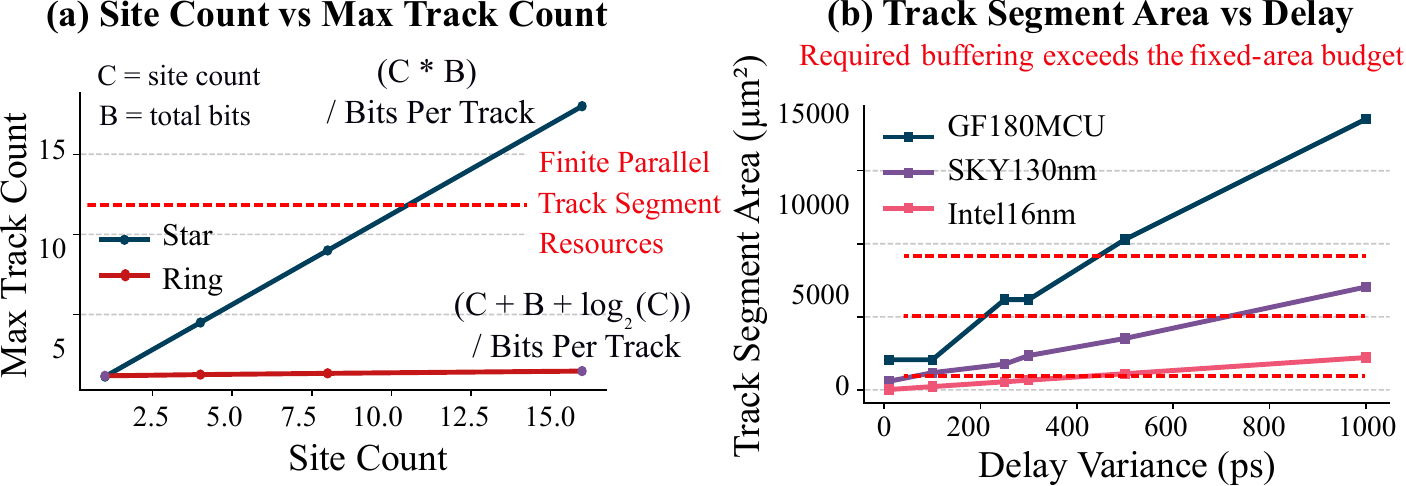}
  \vspace{-0.25in}
  \caption{
  (a)~Star vs.~Ring Interconnect; (b)~Buffer Area Cost}
  \label{fig-star-ring-study-only}
  \vspace{-0.2in}

\end{figure}

Figure~\ref{fig-chipstitch-arch}a shows that the top-level interface is based on a JTAG port, but this is not fundamental and can be replaced with higher-speed links. A global controller serializes JTAG input into packets and injects them into the interconnect. To satisfy the fixed-area constraint, we adopt a single-master, 1D bi-directional mesh network-on-chip with only one active site at a time. This choice eliminates scaling pressure on routing resources. As illustrated in Figure~\ref{fig-star-ring-study-only}a, a star topology requires the controller to drive $N$ independent communication paths, which in the worst case must all be routed through the same narrow channel. This leads to routing demand that scales linearly with $N$, violating the bounded-area requirement. Figure~\ref{fig-star-ring-study-only}b uses rough, approximate fanout-of-4 (FO4) delays in three technologies (180nm, 130nm, 16nm) to show how the general trend of increasing buffer delay needed by the tool to fix timing will take increasing area. Various track widths are annotated with dotted red horizontal lines. Regardless how wide the narrow channel is, required buffering can exceed the fixed-area requirement.
In contrast, a 1D mesh forwards traffic sequentially between neighbor sites. This limits parallelism and path diversity, but ensures that routing resources remain constant regardless of system size. We deliberately avoid closing the topology into a ring to reduce routing, leaving all sites arranged in a single logical line. The interconnect uses single-entry queues, trading performance for lower area by exploiting the non-measurement phases of staging and readback. We avoid deadlock by activating one site at a time and relying on a ready–valid protocol for backpressure. Credit-based flow control was avoided due to area overhead and poor scaling.

Figure~\ref{fig-chipstitch-arch}c shows the host-to-block (h2b) and block-to-host (b2h) messages between the global controller and the sole active site. The bundle is constant-bitwidth regardless of site count.
Figure~\ref{fig-chipstitch-arch}d shows the site station datapath. Registers, queues, and pipeline elements are also fixed-bitwidth and constant-area. We explicitly replace all global buses with locally stored state for a constant overhead per station. For example, an $N$-bit reset signal bundle exceeds track segment area, but a 1-bit reset signal writing a 1-bit reset register per station does not.
We adopt a \IT{global memory-mapped I/O (MMIO) abstraction} (Figure~\ref{fig-chipstitch-arch}b)
with all site control registers in a unified, tessellation-wide address space, configurable power enables and orchestrating data flow.
No component of the architecture scales with number of sites, ensuring compatibility with the narrow interstitial channels imposed by the packing solution.

\subsection{Unbounded Timing VLSI Challenge}

\begin{figure}[!t]
  \centering
  \includegraphics[width=1.0\cw]{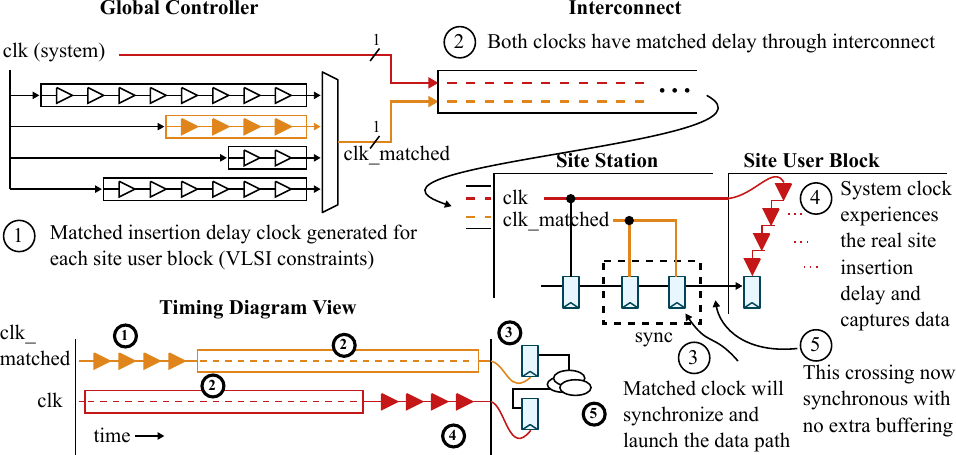}
  \vspace{-0.15in}
  \caption{
  Clock insertion delay-matched \TT{clk\_matched} signal}
  \label{fig-matched-clocking}
  \vspace{-0.1in}

\end{figure}

A key challenge is preventing commercial ASIC tools from fixing timing and violating the fixed-area constraint. Specifically, automated setup-hold fixing through buffer insertion can cause area to grow unpredictably and exceed the narrow interstitial channel budget. Figure~\ref{fig-matched-clocking} shows the boundary between a site station and its user block. Because the site has an independent clock tree with non-zero insertion delay, signals crossing into the site violate STA setup-hold constraints. Tools resolve this with buffering (e.g., on the 65-bit h2b request message), matching the intra-site insertion delay.
\IT{This behavior is fundamentally incompatible with a fixed-area design}. Since the site clock insertion delay is unbounded, the required buffering (and therefore area) also grows unbounded. To address this, \BF{we modify the architecture so that timing is close to correct ``by design'' as seen by tools, requiring no buffering}.


As shown in Figure~\ref{fig-chipstitch-arch}d and Figure~\ref{fig-matched-clocking}, we introduce asynchronous boundaries and a delay-matched clock (\TT{clk\_matched}) for each site. The global controller generates the matched clocks mirroring the site clock tree insertion delays. Both system clock and matched clock propagate delay-matched through the interconnect. The system clock then undergoes the actual site clock insertion delay. The launch and capture events appear aligned under STA, eliminating hold violations by design at all station-to-site crossings and preserving the fixed-area constraint. The synchronizers and single-entry asynchronous FIFOs complete the datapath design.

For similar reasons, we convert global clock into piecewise-segmented, local clock trees to meet fixed track areas (Figure~\ref{fig-eval-piecewise-clock-tree}).


\begin{figure}[t]

  \centering
  \includegraphics[width=0.95\cw]{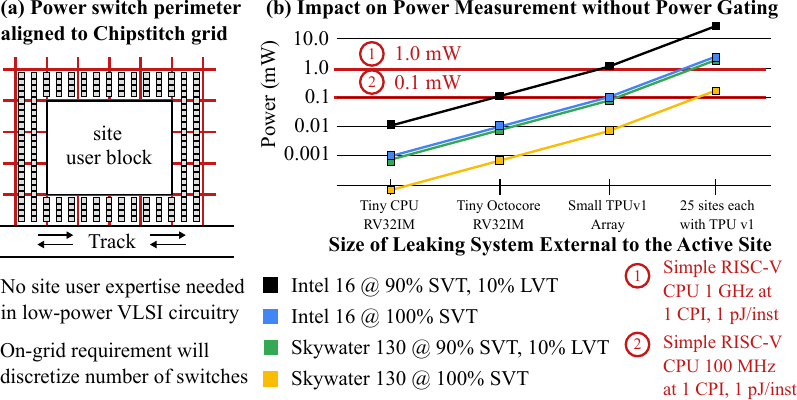}
  \vspace{-0.12in}
  \caption{(a)~Grid-aligned Power Perimeters;
  (b)~Power.}
  \label{fig-power-domains-data}
  \vspace{-0.15in}

\end{figure}


\section{Site Isolation with Power Shutdown}
\label{sec-power}


\begin{figure}[t]

  \centering
  \includegraphics[width=\cw]{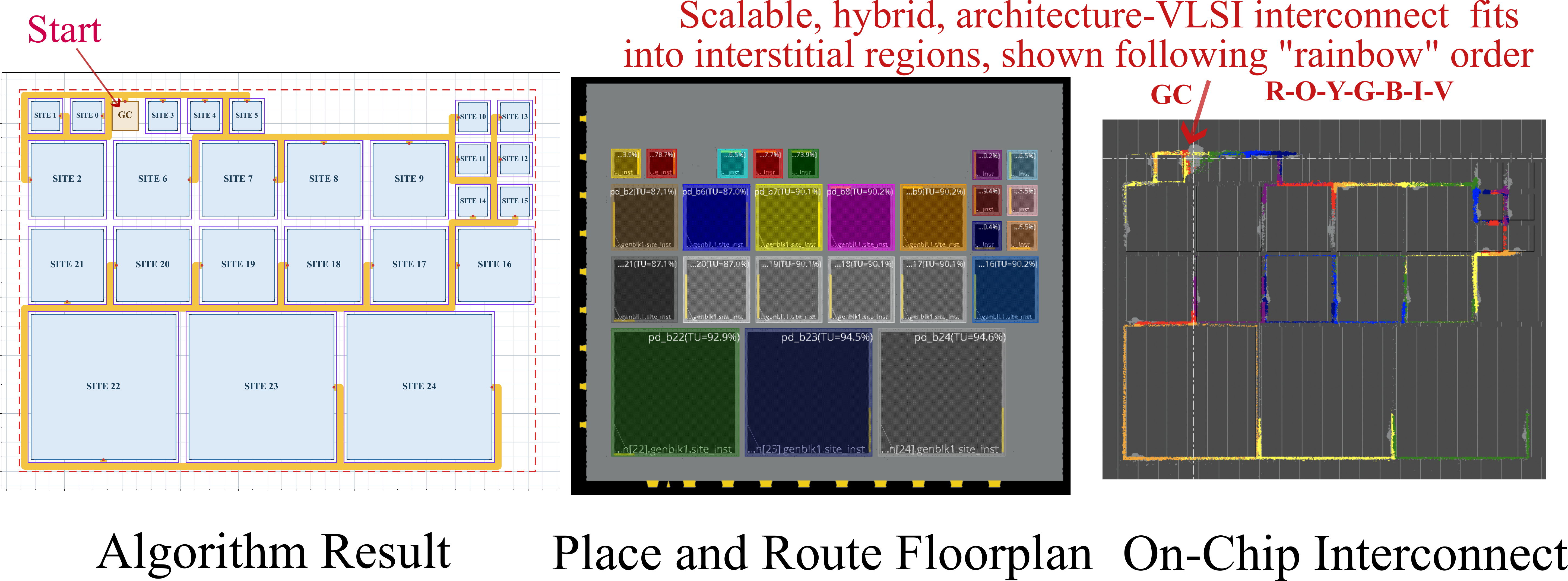}
  \vspace{-0.25in}
  \caption{25-site Chipstitch with Power Domains (colored)}
  \label{fig-eval-chipstitch-overview}
  \vspace{-0.15in}

\end{figure}

Tessellations with a single power supply present a measurement challenge. Inactive sites and system interconnect remain supplied, and leakage (e.g., low-Vt cells) can dominate total power.
Figure~\ref{fig-power-domains-data}b shows post-layout leakage across increasing design scales: a tiny RV32IM CPU, an octocore, a small TPUv1 array, and 25~sites each with a TPUv1. Considering leakage only, and comparing 100\% standard-Vt cells with a 90/10 standard-to-low-Vt mix, full-chip leakage outside the active site grows rapidly with design size. Two horizontal lines show a hypothetical desired measurement of a single RV32IM core operating at 100\,MHz and 1\,GHz (1 CPI, 1\,pJ/inst). On the log-scale y-axis, leakage can exceed and obfuscate this power, making per-site measurement challenging.
To address this, we integrate per-site perimeter-based power domains (see Figure~\ref{fig-power-domains-data}a). Ring-based power gating is standard practice, our contributions are: (1)~embedding grid-aligned perimeters into packing (Section~\ref{sec-algorithms}), which discretizes switch counts; (2)~MMIO-visible configurable control of power shutoff; and (3)~we use high-threshold cells for system design blocks to further reduce the leakage floor.
Figure~\ref{fig-chipstitch-arch}b and~d show the configurable power enable register (en\_pwr\_bar).
Isolation logic is unnecessary. Station selects naturally clamp signals from powered-down sites.
We verify the power shutdown architecture
using Siemens AFS extracted-SPICE simulations in our target technologies
post-signoff for power-off and restore sequences.


\section{Evaluation}
\label{sec-eval}


\begin{figure}[t]

  \centering
  \includegraphics[width=2.3in]{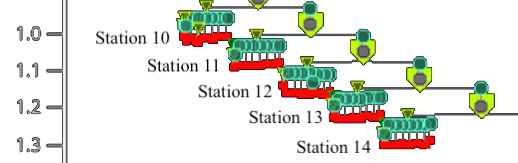}
  \vspace{-0.15in}
  \caption{Piecewise-segmented clock tree visits stations in order, with per-station locally balanced clock trees. Post-layout Innovus meets 2GHz and scales to high site counts.}

  \label{fig-eval-piecewise-clock-tree}
  \vspace{-0.1in}

\end{figure}


\begin{figure}[t]
  \centering
  \includegraphics[width=1.0\cw]{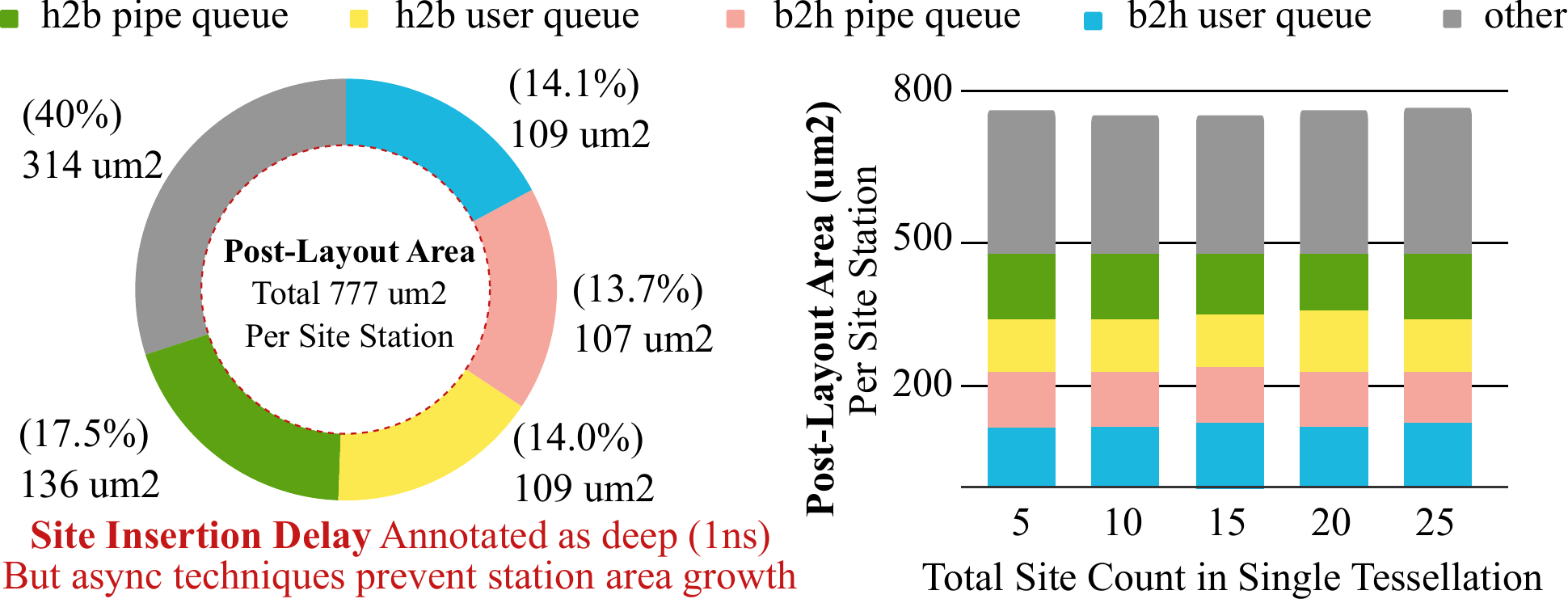}
  \vspace{-0.25in}
  \caption{(left) Post-layout area breakdown in Intel 16nm for single-site station with deep clock insertion delay of 1.0ns, but no area growth. (right) Scalability with tessellation size.}
  \label{fig-results-station-area-scaling}
  \vspace{-0.15in}

\end{figure}

We evaluate a complete silicon query architecture with a 25-site Chipstitch post-layout VLSI implementation in Intel~16nm, as well as Skywater 130nm for comparison to MPW baselines. Our goal is to show that the algorithmic tessellation results transfer realistically into manufacturable layouts, that the scalable interconnect design times and fits within interstitial regions, and that costs do not scale.

\BF{Methodology} -- We produce a signoff-quality post-layout GDS with Cadence Genus and Innovus and complete chip finishing with Synopsys ICV (DRC and LVS-clean). We modified our algorithmic packer to allocate grid squares for mask-alignment foundry cells, tuned site templates via keystone certificate to avoid DRCs (e.g., whitespace, filler gap, endcaps, multi-height cells, special layer patterning) around power perimeters and sites, deployed single-enable daisy power switch cells, custom piecewise CTS, and power-aware VCS gate-level simulation, with limited extracted spice for power switch transients, with two-corner fast-typical STA.

\BF{Conversion to Innovus} -- Figure~\ref{fig-eval-chipstitch-overview} is an algorithm-generated layout for a 25-site tessellation, converted to Innovus floorplan, and fitting the interstitial interconnect into the tessellation (GC start, follow the rainbow color order, end bottom right). Afterwards, sites compose hierarchically (i.e., per-site block-level effort) to sign off.



\begin{figure}[t]
  \centering
  \includegraphics[width=1.01\cw]{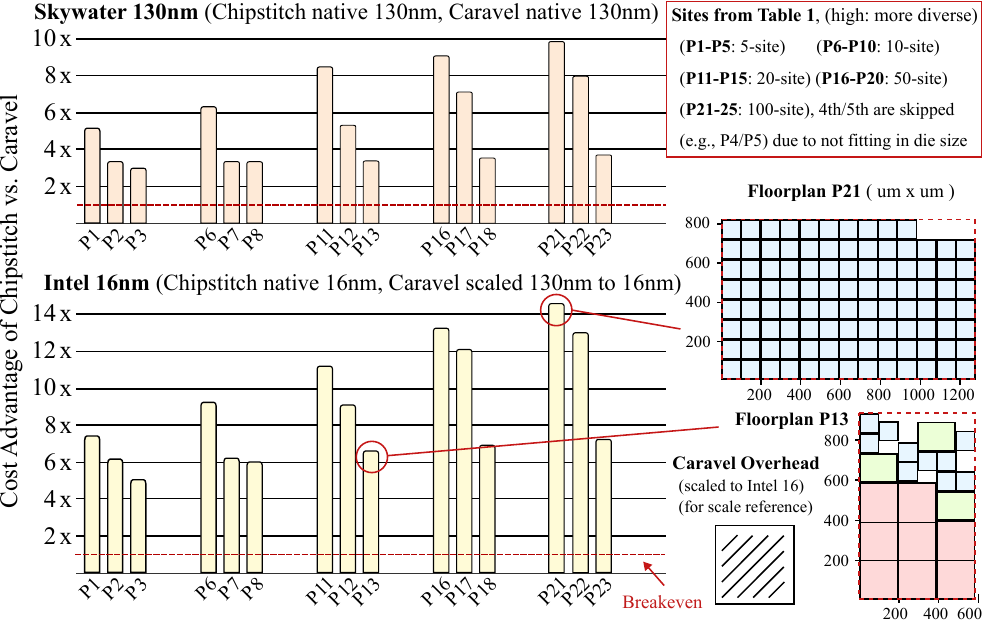}
  \vspace{-0.25in}
  \caption{Total cost advantage of Chipstitch over Caravel~\cite{efabless-web} in cumulative silicon costs for all sites in each set.}
  \label{fig-results-chipstitch-vs-caravel}
  \vspace{-0.15in}

\end{figure}

\BF{Scalable Interconnect Overhead} -- Figure~\ref{fig-results-station-area-scaling} breaks down site station areas showing four single-element queues making up 60\% of the area. The total area is 777 $um^2$. Figure~\ref{fig-results-station-area-scaling} (left) shows how even 1ns intra-site insertion delay does not affect post-layout station area due to our asynchronous crossings and matched clock techniques. Figure~\ref{fig-eval-piecewise-clock-tree} shows a piecewise-segmented clock tree, delivering a 2GHz clock to all sites at fixed area overhead. We also push Intel 16nm post-signoff clock frequency from 100MHz, 250MHz, 1GHz, to 2GHz. GC area increases ($1.00\times$ to $1.45\times$) to push 1GHz timing, compared to site station area ($1.01\times$ to $1.04\times$) up to 2GHz, so the interconnect is not timing-critical. The takeaway is that station overheads do not scale with site count, pushed clock frequencies, or site complexity, confirming our goals from Section~\ref{sec-architecture}.


\BF{Post-Layout Silicon Query} --
We placed a 2$\times$2 systolic array accelerator into the top-right most site in Figure~\ref{fig-eval-chipstitch-overview} (14th of 25 in station visit order) and simulated a complete silicon query within the full-chip tessellation.
From GC, messages and clocks traverse 14~stations under track width of 20um, completing slow data staging (1~Mbps) of 0.5KB matrices into the intra-site register file over the clk to 14th clk\_matched crossing.
System clock is set to 1.75GHz.
Execute phase begins by writing a 1b start flag into an accelerator control register.
The site computes at full speed until done, writing final cycle count (140 cycles) and output data to the intra-site register file.
The experiment completes with a slow read back.
We verify the functional result, satisfying the minimum bar of a silicon receipt. Comparing the silicon query within tessellation vs the same accelerator in isolation, performance matches (140 vs~140 cycles, both 1.75GHz), design area matches (10K~\umsq{}) through identical intra-site floorplanning target, and loop-mode power matches within 0.2\% (1.849mW vs 1.846mW) due to power switch on-leakage, after subtracting idle system and all-site power switch off-leakage.

\BF{Total Cost Advantage} -- Figure~\ref{fig-results-chipstitch-vs-caravel} quantifies cost advantage for Chipstitch over the Efabless Caravel MPW baseline~\cite{efabless-caravel-web, chipfoundry-web} by accumulating total area for each set of sites (P1--P25) including power perimeters, IO pads, system overheads, and manufacturing cells, and computing dollar cost according to Table~\ref{fig-intro}. For Skywater 130nm, we compare Chipstitch and Caravel in the same technology, showing up to a 9.9$\times$ cost advantage for Chipstitch across P-sets. For Intel 16nm, we compare our 16nm Chipstitch layouts against 130nm Caravel (with first-order area scaling to 16nm, Table~\ref{tbl-eval-assumptions}) and multiply both by Intel 16nm cost/\mmsq{}, showing up to 14.5$\times$ cost advantage.
Figure~\ref{fig-results-chipstitch-vs-caravel} draws floorplans P21 and P13 to scale for 16nm, showing various Chipstitch layouts.
The takeaway is that across technologies and sets of sites studied, Chipstitch achieves many integer factors of cost advantage over MPW. This is conservative. Blue sites fit a quad-core RISC-V CPU. Green and red sites are larger. Cost advantage will be higher given more small silicon receipts.

\BF{Single-Receipt Entry Cost} -- Finally, we project first-order entry costs for Intel 16nm silicon receipts using the smallest site studied (i.e., blue square). Consider a 1$\times$1mm, 1\mmsq{} Intel~16nm shuttle. To first order, a 10$\times$10 partitioning provisions 10,000\umsq{} per site, or $\approx$40,000 gate equivalents at FinFET node densities of $\approx$4~GE/\umsq{}. At a shuttle cost of \$20,000/\mmsq{}, this corresponds to $\approx$\$200 per site, each able to fit a quad-core RISC-V CPU. Aside from low cost of entry, Chipstitch shifts the cost of the full-chip delivery away from users to an MPW-style silicon query architecture aggregator service, saving effort, and increasing cost but still leaving a 1--2 orders of magnitude drop in entry cost to an advanced node.

\BF{Overall} -- Chipstitch demonstrates that silicon query architectures can provide research-grade access in a demonstrated FinFET node (e.g., 16nm) with research-grade specifications (e.g., multiple-GHz clocking and isolated single-site power measurement) and intra-die effort (e.g., block-level, not full-chip) for low entry cost.


\begin{table}[t]
  {\centering\ctxcaptionsize

  \caption{Skywater 130nm and Intel 16nm Area Scaling}
  \label{tbl-eval-assumptions}
  \vspace{-0.10in}
  \begin{tabular}{l|l}
    \toprule
    \BF{Caravel~\cite{efabless-caravel-web}} & \BF{Value} \\
    \midrule
    Caravel wrapper dimensions (S130) & 2,920 um x 3,520 um \\
    Caravel user project dimensions (S130) & 2,800 um x 1,760 um \\
    Caravel per-project overhead (S130) & 5,350,400 um$^2$ \\
    Area per gate equivalent (S130) & 11.26 um$^2$ / GE \\
    Area per gate equivalent (I16, ballpark) & 0.33 um$^2$ / GE \\
    \BF{Caravel per-project overhead (I16)} & 156,806 um$^2$ \\
    \bottomrule
  \end{tabular}

  }

  \medskip\ctxcaptionsize
  S130 =~Skywater; I16 =~Intel; Numbers from LEF; GE area is NAND2\_X4
  \vspace{-0.12in}

\end{table}


\section{Related Work}
\label{sec-related}

Our closest related work is MPWs~\cite{kahng-mpw-ispd2004, wu-mpw-todaes2008, wu-mpw-iscas2005, lin-mpw-ieeetase2007, kahng-mpw-tcad2007, kahng-mpw-euromask2007, fang-mpw-rl-integration2023, conway-mosis-mpc-1981, tomovich-mosis-1988, cohen-mosis-1981},
tapeout classes~\cite{tsividis-mit-analog-chip-class-ieeetedu1982, burnett-berkeley-chip-class-iscas2018}, and commercial TinyTapeout~\cite{efabless-tinytapeout-web, chipfoundry-web}. Our work contrasts by requiring both low entry cost and research-grade specs (e.g., TinyTapeout is 50MHz) for silicon receipts.
EDA literature has long studied floorplanning and placement heuristics to minimize
overheads~\cite{wong-floorplan-1986, guo-floorplanning-tcad2001, wang-floorplan-dac1991, chen-floorplan-iccad2025},
joint
floorplanning and routing~\cite{yang-pd-isca2025, liu-synergistic-floorplan-soc2024, yoo-floorplan-tvlsi2009}, and
learning-based approaches~\cite{li-pef-tcad2023, liu-gnn-floorplan-tvlsi2024, xu-goodfloorplan-tcad2022, lin-dreamplace-tcad2020}. Despite synergies, silicon query adds specialized architectural context requiring research.
Interconnect literature studies manycore and accelerator-rich systems~\cite{chang-noc-ieee2001, kim-noc-micro2007, lotfi-noc-micro2012, bakhoda-noc-micro2010, yuan-noc-micro2009, Das-app-aware-noc-ieeemicro2009, Mishra-heter-noc-DAC2013, karkar-noc-ieee2018, duraisamy-noc-vlsi2017, lee-noc-ieee2023,
Zheng-adapt-noc-hpca2021, Zheng-chiplet-noc-DAC2020}.
Our work focuses on low-bandwidth interconnects in fixed limited area.
ARM has low-bandwidth AHB and APB~\cite{arm-amba-ahb2010, arm-amba-apb2010} but they are logical IP abstractions and lack joint architecture-VLSI design considerations.


\section{Conclusion}
\label{sec-conclusion}

Academic and industry chip designers have traditionally chosen between silicon with high entry costs or lacking research-grade specifications, both of which restrict innovation of future hardware-software systems through the credibility afforded by silicon receipts. Chipstitch is a silicon query architecture that breaks past traditional multi-project wafer boundaries and transforms single-die area into a substrate for scalable intra-die silicon receipts, reducing entry costs without sacrificing research-grade specifications. We showed algorithms for repeatable and scalable tessellation, scalable interstitial interconnect design, and power shutdown techniques down to post-layout implementation in both legacy and FinFET nodes. The results suggest that pervasively accessible, low-cost, research-grade silicon receipts can be a reality for future hardware-software innovators and a new research frontier for the chip design community.

\section{Acknowledgements}

This work was supported in part by a gift from Intel Corporation, and the NSF Graduate Research Fellowship Program (GRFP) Award \#2236421. We thank Intel for providing access to the Intel 16 PDK.

\newpage

\bibliographystyle{ACM-Reference-Format}
\bibliography{main}
\end{document}